%
    \input harvmac
%
    \message{[HYPERTEX MODE OFF}

    \def\e@tf@ur#1{}
    \def\hth/#1#2#3#4#5#6#7{{\tt hep-th/#1#2#3#4#5#6#7}}

\def\npb#1#2#3{{Nucl.~Phys.} {\bf B#1} (#2) #3}
\def\prd#1#2#3{{Phys.~Rev.} {\bf D#1} (#2) #3}
\def\plb#1#2#3{{Phys.~Lett.} {\bf #1B} (#2) #3}
\def\phr#1#2#3{{Phys.~Rep.} {\bf #1} (#2) #3}
\def\hpa#1#2#3{{Helv.~Phys.~Acta } {\bf #1} (#2) #3}
\def\ijmpa#1#2#3{{Int.~J.~Mod.~Phys.} {\bf A#1} (#2) #3}
\def\mpa#1#2#3{{Mod.~Phys.~Lett.} {\bf A#1} (#2) #3}

\lref\NAD{
N. Seiberg, {\it Electric-Magnetic Duality in Supersymmetric
Non-Abelian Gauge Theories}, \npb{435}{1995}{129}, \hth/9411149.}

\lref\Seibone{
N. Seiberg, 
{\it Exact Results on the Space of Vacua of 
Four Dimensional SUSY Gauge Theories},
\prd{49}{1994}{6857}, \hth/9402044.}

\lref\unpub{
R.G. Leigh and M.J. Strassler, unpublished.}

\lref\aharony{O. Aharony,
{\it Remarks on Non-Abelian Duality in N=1 Supersymmetric Gauge Theories},
\plb{351}{1995}{220}, \hth/9502013.}

\lref\emop{
R.G. Leigh and M.J. Strassler,
{\it Exactly Marginal Operators and Duality in
Four Dimensional N=1 Supersymmetric Gauge Theory},
\npb{447}{1995}{95}, \hth/9503121.}

\lref\ADS{
I. Affleck, M. Dine and N. Seiberg,
{\it Dynamical Supersymmetry Breaking in Supersymmetric QCD},
\npb{241}{1984}{493};
{\it Dynamical Supersymmetry Breaking in Four Dimensions 
and its Phenomenology}, \npb{256}{1985}{557}.}

\lref\CERN{
D. Amati, K. Konishi, Y. Meurice, G.C. Rossi and G. Veneziano,
\phr{162}{1988}{169}, and references therein.}

\lref\SV{
M.A. Shifman and A.I Vainshtein,
\npb{277}{1986}{456};
\npb{359}{1991}{571}.}

\lref\DDS{
A.C. Davis, M. Dine and N. Seiberg,
\plb{125}{1983}{487}.}

\lref\NSVZ{
V.A. Novikov, M.A. Shifman, A.I. Vainshtein and V.I. Zakharov,
\npb{260}{1985}{157}.}

\lref\tH{
G. 't Hooft,
any paper.}

\lref\intpoul{
K. Intriligator and P. Pouliot, 
{\it Exact Superpotentials, Quantum Vacua and Duality in Supersymmetric 
$Sp(N_c)$ Gauge Theories}, \plb{353}{1995}{471}, \hth/9505006.}

\lref\kinsso{K. Intriligator and N. Seiberg,
{\it Duality, Monopoles, Dyons, Confinement and Oblique Confinement in 
Supersymmetric $SO(N_c)$ Gauge Theories},
\npb{444}{1995}{125}, \hth/9503179.}

\lref\kinsrev{K. Intriligator and N. Seiberg,
{\it Lectures on Supersymmetric Gauge Theories and Electric-Magnetic
 Duality,} Nucl.~Phys.~Proc.~Suppl. {\bf 45BC} (1996) 1,
 \hth/9509066.}

\lref\kutsch{
D. Kutasov,
{\it A Comment on Duality in $N=1$ Supersymmetric Non-Abelian Gauge Theories},
\plb{351}{1995}{230}, \hth/9503086; 
D. Kutasov and A. Schwimmer,
{\it On Duality in Supersymmetric Yang-Mills Theory},
\plb{354}{1995}{315}, \hth/9505004; 
D. Kutasov, A. Schwimmer and N. Seiberg,
{\it Chiral Rings, Singularity Theory and Electric-Magnetic Duality},
\npb{459}{1996}{455}, \hth/9510222.}

\lref\ilslist{K. Intriligator, {\it New RG Fixed Points and
Duality in Supersymmetric $Sp(N_c)$ and $SO(N_c)$ Gauge
Theories}, RU--95--27, \npb{448}{1995}{187}, \hth/9505051; 
R.G. Leigh and M.J. Strassler, {\it Duality of
$Sp(2N_c)$ and $SO(N_c)$ Supersymmetric Gauge Theories with
Adjoint Matter}, \plb{356}{1995}{492}, \hth/9505088;
K. Intriligator, R.G. Leigh and M.J. Strassler, 
{\it New Examples of Duality in Chiral and Non-Chiral Supersymmetric
Gauge Theories}, \npb{456}{1995}567, \hth/9506148.}

\lref\spinduals{P. Pouliot, 
{\it Chiral Duals of Non-Chiral SUSY Gauge Theories},
\plb{359}{1995}{108}, \hth/9507018;
P. Pouliot and M.J. Strassler,
{\it A Chiral $SU(N)$ Gauge Theory and its Non-Chiral $Spin(8)$ Dual},
\plb{370}{1996}{76}, \hth/9510228. }

\lref\threed{
K. Intriligator and N. Seiberg, 
{\it Mirror Symmetry in Three Dimensional Gauge Theories},
\hth/9607207.}

\lref\DSBaryons{
C.~Cs\'{a}ki, L.~Randall and W.~Skiba, 
{\it More Dynamical Supersymmetry Breaking},
MIT-CTP-2532, \hth/9605108; C.~Cs\'{a}ki, R.G.~Leigh, 
L.~Randall and W.~Skiba,
{\it Supersymmetry Breaking Through Confining and Dual Gauge Theory Dynamics}, 
\plb{387}{1996}{791}, \hth/9607021.
}

\def\bib#1#2#3#4#5{#1 {#2{#3}{19#5}{#4}}}
\lref\suthreefinite{
\bib{A. Parkes and P. West,}{\plb}{130}{99}{84};
\bib{P. West,}{\plb}{137}{371}{84};
\bib{D.R.T. Jones and L. Mezincescu,}{\plb}{138}{293}{84};
\bib{S. Hamidi, J. Patera and J. Schwarz,}{\plb}{141}{349}{84};
\bib{S. Hamidi and J. Schwarz,}{\plb}{147}{301}{84}.
}

\lref\allorders{
\bib{O. Piguet and K. Sibold,}{\ijmpa}{1}{913}{86};
\bib{}{\plb}{177}{373}{86};
\bib{A.V. Ermushev, D.I. Kazakov and O.V. Tarasov,}{\npb}{281}{72}{87};
C. Lucchesi, O. Piguet and K. Sibold,
{\it Conf. on Differential Geometrical Methods in Theoretical Physics},
Como, 1987;
\bib{}{\hpa}{61}{321}{88};
\bib{D.I. Kazakov,}{\mpa}{2}{663}{87};
{\it Ninth Dubna Conf. on the Problems of Quantum Field Theory}, Dubna,
 1990;
\bib{X.-D. Jiang and X.-J. Zhou,}{\prd}{42}{2109}{90}.
}

\def\qt{\ot{q}}

\def\rarr{\rightarrow}
\def\frac#1#2{{#1\over #2}}
\def\del{\partial}

\def\al{\alpha}
\def\bt{\beta}
\def\lam{h}
\def\lt{\ot{\lam}}
\def\Lam{\Lambda}
\def\LamD{\tilde\Lambda}
\def\Lb{\Lam^b}

\def\LbD{\tilde{\Lam}^{\tilde{b}}}
\def\LbDL{\tilde{\Lam}_L^{\tilde{b}_L}}
\def\eps{\epsilon}
\def\hlf{{1\over2}}
\def\ot#1{\tilde{#1}}
\def\h{\hat}
\def\vev#1{\langle #1\rangle}
\def\su#1{${SU(#1)}$}
\def\sunc{\su{N_c}}
\def\ncd{{N_f-N_c}}
\def\sunfc{\su{\ncd}}
\def\sunfLRU{$SU(N_f)_L\times SU(N_f)_R\times U(1)_B$}
\def\suii{\su{2}}
\def\sunnf{\su{2N_f}}
\def\fnc{${\rm\bf N}_c$}
\def\anc{$\overline{\rm\bf N}_c$}
\def\nf{{N_f}}
\def\nc{{N_c}}

\def\Miti{$M_{u}^{\; r}$}
\def\Mitim{M_{u}^{\; r}}

\def\fQn#1{Q^{#1}}
\def\aQn#1{\ot{Q}_{#1}}
\def\fqn#1{q_{#1}}
\def\aqn#1{\ot{q}^{#1}}

\def\wdual{$W=\mu^{-1}\fqn{}\cdot M\cdot\aqn{}$}
\def\wtree{W_{{\rm tree}}}

\def\lis{\lam_{r_1\ldots r_{N_c}}}
\def\ltis{\ot{\lam}^{u_1\ldots u_{N_c}}}

\def\Bis{B^{r_1\ldots r_{N_c}}}
\def\Btis{\ot{B}_{u_1\ldots u_{N_c}}}

\def\Qis{\epsilon_{\al_1\ldots\al_{N_c}}\;
\fQn{\al_1r_1}\ldots\fQn{\al_{N_c}r_{N_c}}}

\def\qis{{1\over (N_f-N_c)!} \epsilon^{r_1\ldots r_{N_f}}
\epsilon^{\bt_1\ldots\bt_{(N_f-N_c)}}\;
\fqn{\bt_1 r_{N_c+1}}\ldots \fqn{\bt_{(N_f-N_c)}r_{N_f}}}

\def\Title#1#2{\nopagenumbers\abstractfont\hsize=\hstitle\rightline{#1}%
\vskip .45in\centerline{\titlefont #2}\abstractfont\vskip .5in\pageno=0}

\Title{
{
 \vbox{	\hbox{hep-th/9611020}
	\hbox{RU--96--101}
	\hbox{ILL--(TH)--96--11}
	\hbox{IASSNS--HEP--96/109}
}}}
{\hsize =\hstitle \vbox{
	\centerline{Accidental Symmetries and N=1 Duality}
	\bigskip
	\centerline{in Supersymmetric Gauge Theory}
}}

\centerline{Robert G. Leigh\foot{e-mail: {\tt rgleigh@uiuc.edu}}}
{\it 	\centerline{Department of Physics}
	\centerline{University of Illinois at Urbana-Champaign}
	\centerline{Urbana, IL 61801, USA}
} 
\vglue .5cm
\centerline {Matthew J. Strassler\foot{e-mail: {\tt strasslr@sns.ias.edu}}}
{\it 	\centerline{School of Natural Sciences}
	\centerline{Institute for Advanced Study}
	\centerline{Princeton, NJ 08540, USA}
}
 
\bigskip\bigskip
\baselineskip14pt\noindent
We note that the accidental symmetries which are present in some
examples of duality imply the existence of continuously infinite sets
of theories with the same infrared behavior.  These sets interpolate
between theories of different flavors and colors; the change in color
and flavor is compensated by interactions (often non-perturbative)
induced by operators in the superpotential.  As an example we study
the behavior of $SU(2)$ gauge theories with $2\nf$ doublets; these are
dual to $SU(N_f-2)$ gauge theories whose ultraviolet flavor symmetry
is \sunfLRU\ but whose flavor symmetry is \sunnf\ in the infrared.
The infrared \sunnf\ flavor symmetry is implemented in the ultraviolet as a
non-trivial transformation on the Lagrangian and matter content of the
magnetic theory, involving (generally non-renormalizable) baryon operators and
non-perturbative dynamics.  We discuss various implications of this
fact, including possible new chiral fixed points and interesting
examples of dangerously irrelevant operators.

\Date{11/96 (12/94)}
\baselineskip16pt

\newsec{Introduction}
\seclab{\intro}

Recent work has uncovered many interesting aspects of four-dimensional
N=1 supersymmetric gauge theories. Foremost is the electric-magnetic
duality discovered by Seiberg \NAD. Completely different, yet
equivalent, descriptions may exist for a given field theory. It is
apparent that formerly sacred principles must be reconsidered; for
example the gauge group does not so much {\it define} a model, as give
a (perhaps weakly-coupled) description of that model.  Global symmetry
however is believed to encode aspects of the physical content of the
theory. Many theories which are dual to one another share the same
global symmetries in the ultraviolet (UV) as well as in the infrared
(IR).  However, the matching of global symmetries need not occur at
the level of the perturbative definition of two dual theories; rather,
it need only occur in the far IR.  Many models exist for which the
global symmetries of the UV descriptions are different.  

In this paper, we explore some consequences of this fact.  We consider the
simple example of $SU(2)$ gauge theories with $2\nf$ doublets,
which have a dual description in terms of $SU(\nf-2)$ gauge theories that
have $\nf$ flavors and a smaller global symmetry.  We
show how the latter theories have an 
``accidentally'' enhanced global symmetry in the IR,
through a mixture of perturbative and
non-perturbative physics.  Using this we uncover non-trivial continuously
infinite classes of theories which all flow to the same IR theory.  We
also discuss a number of interesting issues which are raised
by our discussion, including examples of
dangerously irrelevant operators and new chiral fixed points.

\newsec{Baryons in $SU(N_c)$ Gauge Theories}
\seclab{\general}

We begin with some introductory material on N=1 supersymmetric
$SU(N_c)$ gauge theory, paying particular attention to the properties
of baryon operators under the duality mapping.  The dynamics studied
here will be used in Sec.~3.

We consider an \sunc\ supersymmetric gauge theory with $N_f$ flavors
of chiral superfields $\fQn{\al r}$ in the \fnc\ and $\aQn{\al u}$ in
the \anc\ of \sunc.  We will refer to this as the electric theory.
This model has an \sunfLRU\ global symmetry.\foot{Throughout the paper
we use the letters $\al,\bt$ for color indices, $r,s,\ldots$ for
$SU(N_f)_L$ and $u,v,\ldots$ for $SU(N_f)_R$.}  According to the
duality proposed by Seiberg \NAD, the magnetic theory is \sunfc\ with
$N_f$ flavors $\fqn{r}^\al$ and $\aqn{u}_\al$,
along with gauge singlet superfields \Miti\ and a superpotential \wdual;
here $\mu$ is a scale factor needed to match the dimensionful
quantities of the magnetic theory to those of the electric theory
\refs{\unpub,\kinsso,\kinsrev,\kutsch}.\foot{Specifically, the 
holomorphic dynamical scales $\Lam$ and $\LamD$ of the electric and 
magnetic theories are related by $\Lb\LbD=(-)^{\nf-\nc}\mu^\nf$, where 
$b=3\nc-\nf$ and
$\tilde b = \tilde\nc -\nf=2\nf-3\nc$ are the coefficients of the
one-loop beta function in the two theories.}  The magnetic theory also
possesses a global \sunfLRU\ symmetry; note that the fields $\fQn{}$
and $\fqn{}$ transform differently under this symmetry.  If the
electric gauge group is \suii, then the flavor symmetry is enlarged to
\sunnf.  However, the flavor symmetry in its dual remains \sunfLRU;
the full \sunnf\ symmetry is realized only in the IR as
an accidental symmetry.

The gauge-invariant chiral operators of the electric theory are mapped
under the duality to operators in the magnetic theory. In Ref.~\NAD,
it was checked that perturbations involving superpotential
deformations by and expectation values for meson operators preserve
the duality. It may be checked that the same is
true for baryon operators, as was done in part in \refs{\aharony}.

We will refer to the invariant operators of the theory in the
following way.  The mesons \Miti\ of the theory are $M(Q)\equiv
Q\tilde Q$ in the electric theory and are fundamental fields $M$ in
the magnetic dual. The baryons $B^{i_1\dots i_{N_c}}$ are
\eqn\baryelec
{ \Bis(Q)= \Qis}
in the electric theory and
\eqn\barymag
{\Bis(q)=  \sqrt{-\Lb\over (-\mu)^{\nf-\nc}}\qis}
in the magnetic theory \kinsrev.
Analogous formulas hold for the antibaryons. 

We now consider a general $SU(\nc)$ theory with $\nf\geq\nc+2$ flavors
with a superpotential which contains baryon operators, and study its
flow and that of its dual when mass terms are added or symmetries are
broken.  (Some of this discussion also appears in Ref.~\aharony.)  We
begin by comparing the flat directions in the two theories.

Seiberg \NAD\ showed that the flat directions of the
theory and its dual match in the absence of a superpotential. Some
of these directions are lifted by the superpotential
\eqn\barypert{
\wtree = {1\over\nc!}\left[\lis\Bis(Q) + \ltis\Btis(\aQn{})\right].}
{}From the F-flatness conditions $\vev{\frac{\del W}{\del Q^r}}=0$ 
one finds
\eqn\FflatGen{
\vev{Q^s\frac{\del W}{\del Q^r}} \propto
\vev{h_{r r_2\dots r_\nc} B^{s r_2\dots r_\nc}} = 0
}
for all $r,s$, and similar conditions on $\ot h \ot B$.

In the magnetic theory 
\eqn\barypertd{\wtree = \mu^{-1} \Mitim q_r\qt^u +
{1\over\nc!}\left[\lis\Bis(q) + \ltis\Btis(\aqn{})\right]
}
the flatness conditions are
\eqn\FflatGenD{
\eqalign{
\vev{\frac{\del W}{\del q_r}} =
\vev{{1\over\nc!}\lis \frac{\del}{\del q_{r}} B^{r_1\dots r_\nc}(q)
                + {1\over\mu}\Mitim \qt^u} = 0 \ \cr
\vev{{1\over\nc!}\ltis \frac{\del}{\del \qt^u} \ot B_{u_1 \dots u_\nc}(\qt)
                + {1\over\mu}\Mitim q_r} = 0 \cr
\vev{q_r\qt^u}=0 \ .
}
}
Multiplying the first equation of
\FflatGenD\ by $q_s$ and applying the third equation,
we find the condition \FflatGen.  The condition on the antibaryon
is similarly recovered.

There are also restrictions on the expectation values of the operators
$M^r_u$.  We consider the simplest case, when all components of $h,\ot
h$ are zero except $h_{12\dots\nc}$ (and its permutations).  The
F-flatness conditions, which force the $\nc\times\nc$ matrix $Q^{\al
r}$ to have rank $\nc-2$ or less, impose a similar constraint on the
matrix $M^{\bar r}_u(Q)$, where $\bar r = 1,\cdots,\nc$ and
$u=1,\cdots,\nf.$ To see this in the magnetic theory, let us first
note that there is no constraint on $M^r_u$ from \FflatGenD\ if all
$q$ and $\qt$ are zero.  However, as in \NAD, a dynamical
superpotential is generated if $M$ is rank $\nc-1$.  For simplicity
let $M$ be diagonal with only $\vev{M^1_1}, \vev{M^2_2},\cdots
\vev{M^{\nc-1}_{\nc-1}}$ non-vanishing.  
Such a choice gives mass to the fields $q_r, \qt^u$,
$r,u=1,\cdots,\nc-1$.  When they are integrated out the theory
confines \refs{\Seibone,\NAD}, and the superpotential is a function of
the singlet superfields $M^{\h r}_{\h u}$, $N^{\h u}_{\h r}=\fqn{\h
r}\aqn{\h u}$, $B_L^{\h r}\propto B^{12\dots(\nc-1)\h r}(q)$ and $\ot
B_{L\h u}\propto \ot B_{12\dots(\nc-1)\h u}(\ot q)$ , where $\h r, \h
u = \nc,\nc+1,\dots,\nf$.
\eqn\Wdynbloc{\eqalign{
W(M,N,B,\ot B) =& \mu^{-1}M^{\h r}_{\h u} N^{\h u}_{\h r} + 
(h_L)_{\nc} B_L^\nc  
- \sum_{a=1}^{\nc-1}{N^{\h u}_{\h r}M^{\h r}_{a}M^{a}_{\h u}\over \mu M^a_a}
\cr & -  {\det (N^{\h u}_{\h r})\over\LbDL} -
{B_L^{\h r} N^{\h u}_{\h r} {\ot B}_{L\h u}\over\mu} 
\ .\cr}
 } 
(The last two terms are due to strong coupling effects \Seibone\
while the term $N^{\h u}_{\h r}M^{\h r}_{a}M^{a}_{\h u}/\mu M^a_a$ is
induced at tree level by integrating out the massive $q_r, \qt^u$
fields.)  Here $(h_L)_{\nc}={\vev{Q^1Q^2\dots
Q^{\nc-1}}}h_{12\dots\nc} =\sqrt{\vev{M^1_1M^2_2\dots
M^{\nc-1}_{\nc-1}}}h_{12\dots\nc}$ is the low-energy baryon coupling,
and the formula $\LbDL=\vev{M^1_1M^2_2\dots
M^{\nc-1}_{\nc-1}}\mu^{1-\nc}\LbD$ relates the dynamical scales of the
high- and low-energy magnetic theories.  The conditions
\eqn\cantbe{\eqalign{
\vev{\frac{\del W}{\del M^\nc_{\h u}}} \propto \vev{N^{\h u}_\nc} = 0 \cr
\vev{\frac{\del W}{\del B_L^\nc}}
  = (h_L)_\nc- \vev{N_{\nc}^{\h u}{\ot B}_{L\h u}}\mu^{-1} = 0 }}
are inconsistent.  This confirms that those flat directions with ${\rm
rank}\left(\vev{M^{\bar r}_u}\right)>\nc-2$ are lifted by confinement
effects.  We will repeatedly encounter similar situations in which
flat directions are ``dynamically blocked'' by the baryon operators in
the superpotential.

We have shown here that the submatrix $M^{\bar r}_u$ cannot have rank
greater than $\nc-2$.  It is easy to see that if we had taken $M$ of
rank $\nc-1$ but with $M^{\bar r}_u$ of lesser rank, we would
not have had the second of equation \cantbe, and we would have found
consistent solutions to the F-flatness conditions.  When we study the
\suii\ case below, we will exhibit this explicitly.

\newsec{Accidental Symmetries for Duals of \su{2}\ Gauge Theories}
\subseclab{\sutwo}

When the gauge group is \su{2}, both mesons and baryons are quadratic
in the underlying fields. Since the fundamental of \su{2}\ is
pseudoreal, there is no invariant distinction between $\fQn{}$ and
$\aQn{}$; as a result the flavor symmetry is enhanced to \sunnf\ and
baryon perturbations \barypert\ are simply mass terms.  We will
retain, for our present purposes, the usual \sunc\ labelling
convention, which makes an artificial distinction between $\fQn{}$ and
$\aQn{}$.  The gauge invariant bilinears, which transform as an
antisymmetric tensor $V(Q)$ of $SU(2\nf)$, then break up into
``mesons'' and ``baryons'' as follows:
\eqn\sutwoV{V(Q)=\pmatrix{ B(Q)&M(Q)\cr -M^T(Q)&\tilde B(\ot Q)}.}

 As described in Ref.~\NAD, for $\nf>3$ the theory has a magnetic
description using a gauge group \su{\nf-2}, $\nf$ flavors and a
superpotential \wdual, where \Miti\ are gauge singlets which map to
$Q^r\tilde Q_u$.  Clearly this theory, at least in the UV, does not
possess the full \sunnf\ symmetry.\foot{ This theory has an alternate
dual description which does explicitly preserve \su{2\nf}\ invariance.
The singlet fields and the assignment of charges to the $q$ fields are
different in this case, and the duality is a member of the $Sp$ series
of Refs.~\refs{\NAD,\intpoul}.} However, in the IR theory (which is in
a non-Abelian Coulomb or free magnetic phase, depending on $\nf$), the
operators of interest are the gauge invariants $M$, $B(q)$, $\tilde
B(\tilde q)$ and $N=q\aqn{}$. The latter is a redundant operator, and
the remaining invariants transform in a representation of \sunnf, as
in Eq.~\sutwoV.

We consider adding to the \su{2}\ theory a superpotential of
the form:
\eqn\sutwoW{
\wtree=m_r^u Q^r\tilde Q_u +\hlf\lam_{rs}B^{rs}(Q)+
\hlf\lt^{uv}\tilde B_{uv}(\aQn{}).
}
The F-term conditions arising from \sutwoW\ are
\eqn\fterm{\eqalign{
\vev{\frac{\del W}{ \del\fQn{\al r}}}=&
\vev{m_r^u\aQn{\al u}+\eps_{\al\bt}\lam_{rs}\fQn{\bt s}}=0\cr
\vev{\frac{\del W}{ \del\aQn{\al u}}}=&
\vev{m_r^u\fQn{\al r}+\eps^{\al\bt}\lt^{uv}\aQn{\bt v}}=0
}} 
If we perform a rotation in \sunnf\ on this theory, the various
terms in \sutwoW\ will simply rotate into one another, and the physics
is unchanged.  However, such a transformation has a non-trivial effect
on the magnetic theory away from the IR fixed point.  We now explore
this effect, looking at the simplest cases of $m$ of rank one or $h$ of
rank one, and the interpolation between them.

\subsec{$m\neq 0$}
\subseclab{\twomass}

We consider first the case \NAD\ where $m$ has rank one and $\lam=\lt=0$:
\eqn\sutwoWm{
\wtree=m \fQn{1}\aQn{1} \ .
} 
For large values of $m$, the fields $\fQn{1},\aQn{1}$ are massive
and can be integrated out. For $\nf>4$, this procedure sets
$\vev{\fQn{1}}=\vev{\aQn{1}}=0$ and leads to a low-energy \su{2}\
gauge theory with $\nf-1$ flavors and no superpotential.

In the magnetic $SU(\nf-2)$ gauge theory, the inclusion of \sutwoWm\
results in a superpotential
\eqn\dualWm{
\tilde\wtree=\frac{1}{\mu}\fqn{r} \Mitim\aqn{u}+m M^1_1\ .
}
Here the quarks $\fqn{1}$ and $\aqn{1}$ have a vacuum
expectation value given by
\eqn\massbrk{
\vev{\fqn{1}\cdot\aqn{1}}=-m\mu
}
{}from the $F$-flatness condition for $M$.
This  breaks the gauge symmetry down to \su{\nf-3}
with $\nf-1$ flavors, which is indeed dual \NAD\ to
the low-energy \su{2}\ electric theory with $\nf-1$ flavors.  For
$\nf-1>5$ the duality is manifested through the flow of
the magnetic theory to the free electric theory in the IR;
for $\nf-1=4$ or $5$ both theories flow to the same interacting
IR fixed point.

\subsec{$\lam\neq 0$}

We now consider the case where only $\lam_{12}\equiv\lam\neq0$. From
the point of view of the electric theory, this is a trivial flavor
rotation $\aQn{1}\rarr\fQn{2}$ of the case studied above.  The
equations \fterm\ reduce to $\vev{\fQn{\al 1}}=\vev{\fQn{\al 2}}=0$,
and thus
$\vev{M^1_u(Q)}=\vev{M^2_u(Q)}=\vev{B^{1r}(Q)}=\vev{B^{2r}(Q)}=0$,
where $r=3,\dots,\nf$ and $u=1,\dots,\nf$.
Taking into account the D-flatness conditions, we find that \Miti$(Q)$
can be at most rank 2, and $B(Q)$, $\ot B(\ot Q)$ at most rank 1,
with the constraint
\eqn\hconstr{M^r_u(Q) M^s_v(Q) - M^r_v(Q) M^s_u(Q) =
B^{rs}(Q)\ot{B}_{uv}(\ot Q)\ .}  
We wish to show now that the same
flat directions are found in the magnetic theory.

The magnetic theory is \su{\nf-2}\ with $\nf$ flavors and a superpotential
\eqn\nftwosupd{
\wtree=\frac{1}{\mu}\fqn{r}\Mitim\aqn{u}+\lam B^{12}(q).
}
Although $h$ acts as a mass term in the electric theory, the physics in
the magnetic dual is not that of symmetry breaking, in contrast
to the case of non-zero $m$.  Instead,
the analysis of section \general\ applies here and one
easily finds the condition
$\vev{B^{1r}}=\vev{B^{2r}}=0$ as in \FflatGenD.  Furthermore,
as in \Wdynbloc--\cantbe,
the flat directions labelled by $M^1_u$ and $M^2_u$ are dynamically
blocked.  We now show that the remaining
part of $M$ must be rank 2 and satisfies the correct constraint \hconstr.

Consider the direction where $M^3_3$
is non-zero. This vacuum expectation value gives mass to the fields
$\fqn{3},\aqn{3}$, which should be integrated
out at low energy scales, leading to a low-energy theory with
one more flavor than color.  This theory confines and generates a dynamical
superpotential for the confined fields ${N}_{\h r}^{\h u}=\fqn{\h r}\aqn{\h
u}$
and ${B_L}^{\h r}\propto B^{3\h r}(q)$, 
$\ot{B}_{L\h u}\propto\ot{B}_{3\h u}(\ot q)$
(here, $\h r=1,2,4\dots\nf$) 
\eqn\lowdynsupd{
\tilde W_L=\mu^{-1}M^{\h r}_{\h u} {N}^{\h u}_{\h r}
-{\mu\over M^3_3}
{\det{N}^{\h u}_{\h r}\over\LbD}  -
{{B_L}^{\h r}{N}_{\h r}^{\h u}
{\ot B}_{L\h u}\over\mu}
-{{N}_{\h r}^{\h u}M^{\h r}_3 M^3_{\h u}\over \mu M^3_3} \ . }
Compare this with \Wdynbloc, noting the absence of the $hB$ term.  The
flatness conditions have solutions with $N=0$ and $M^{\h r}_{\h
u}=(B^{3\h r}{\tilde B}_{3\h u}+M^{\h r}_3M_{\h u}^3)/M^3_3$, which is
one of the constraints \hconstr\ of the electric theory. A similar
analysis holds if $M$ is rank 2, for which the number of flavors and
colors are equal. If we take $M_3^3$, $M^4_4$ nonzero, then instead of
a dynamical superpotential there is a constraint \Seibone
\eqn\constrdyn{
\det ({N}_{\h r}^{\h u})+ {(-\mu)^{\nf-2}\over\Lb}{B}
{\ot B} = \ot\Lam_L^{\ot b_L} = M^3_3 M^4_4 {\LbD\over \mu^2}} 
where $\h
r=1,2,5\dots\nf$, $B=B^{34}$ and $\ot{B} = \ot{B}_{34}$. 
Here the only solution to the
flatness conditions is $ M^{\h r}_{\h u} ={N}_{\h r}^{\h u}=0$,
so Eq.~\constrdyn\ satisfies the constraint \hconstr.  
Finally, for $M$ of higher rank, the theory has fewer flavors than
colors, and all flat directions are lifted by a dynamical
superpotential \refs{\ADS,\Seibone}. One can also show that the baryons
are at most rank one.  Thus, the flat directions in the two theories
are indeed the same.

\subsec{$h,m\neq0$}

We may interpolate between the two cases studied above by letting both
$m_1^1=m$ and $\lam^{12}=\lam$ be non-zero.  In the electric \suii\
theory, a single flavor becomes massive.  The requirements of
Eq.~\fterm\ imply $M^1_j(Q)=B^{1j}(Q)=0$ and
\eqn\MBmix{
\eqalign{
\vev{mM^r_1(Q) + \lam B^{r2}(Q)}=0 \ ,\cr
\vev{m\ot B_{1u}(Q) + \lam M^2_u(Q)} =0 \ ;
}
}
thus these operators mix through perturbative dynamics.

In the magnetic theory, which is broken to \su{\nf-3}, Eqs.~\MBmix\
are also satisfied.  The first equation can be seen immediately by
considering the F-flatness conditions. The second equation arises in a
more interesting way through non-perturbative effects.  One way to see
this is to take $M^2_2\neq 0$; then $\fqn{2}$ and $\aqn{2}$ are
massive and should be integrated out, resulting in a dynamically
generated superpotential 
\eqn\wdynh{
\ot W_L = m \h{M}_1^1 + \mu^{-1}\h{M}\h{N} 
-{1\over \mu M^2_2}\h{N}_{\h r}^{\h u}
M^{\h r}_2M^2_{\h u} + \lam \h{B}^{1}
-\frac{\det \h{N}}{\LbD M^2_2} -
{\h{B}^{r}\h{N}^u_r\h{\ot{B}}_u\over\mu M^2_2}
}
where hatted operators have indices running over $1,3,4,\ldots,N_f$
and where $\h{B}^{r}=B^{r2}$.  The F-flatness conditions
imply $N^1_1 = -m\mu$; from
$\del{W}/\del{\h{B}^1}$ we find that
\eqn\fflatcon{
\lam M^2_2+ m \ot{B}_{12} = 0
}
Thus, in the magnetic theory, the
mixing of the relevant operator \Miti\ and the
IR-relevant operator $\tilde B(q)$ arises through a combination
of perturbative and non-perturbative dynamics.

To summarize, as we rotate from a theory with $m\neq 0$ to one with
$h\neq 0$ (a trivial symmetry operation in the electric \suii\
theory), we find that the magnetic theory changes in an intricate
way. From $h=0$, where it is an \su{\nf-3}\ gauge theory with $\nf-1$
flavors and no special dynamics, the theory changes to one in which
singlet meson and composite baryon operators mix dynamically, until
finally for $m=0$ the \su{\nf-2}\ symmetry is restored along with all
$\nf$\ flavors, and with gauge dynamics and the superpotential
$\wtree=hB(q)$ restricting the flat directions.  Remarkably, this
continuous transformation of the UV theory does not in any way change
the low-energy dynamics, on which it is realized as a symmetry
transformation of the low-energy fields.

The generalization to mass matrices and $\lam$ matrices of
higher rank is straightforward.

\subsec{Symmetry Breaking in the $SU(2)$ Theory}

It is also interesting to consider the effect of giving a baryon
operator a vacuum expectation value. Let us first review the effect of
$\vev{\fQn{1}\aQn{1}}\neq 0$ which was studied in detail in Ref.~\NAD.
The gauge symmetry is broken completely; the D-flatness conditions
require that \Miti\ have rank at most 2, in which case (after using
the remaining flavor symmetry) we can write a constraint
\eqn\SBconstr{
M^1_1(Q) M^2_2(Q) - M^1_2(Q)M^2_1(Q) = B^{12}(Q)\ot{B}_{12}(\ot Q) \ .
}
With a general mass matrix
$\h{m}_{\h r}^{\h u}Q^{\h r}\ot{Q}_{\h u}$ for the other fields
instanton dynamics generates a superpotential
\eqn\SBvev{
W_L=\Lb \frac{\det \h m}{\fQn{1}\aQn{1}}
}
The magnetic theory with $\vev{M^1_1}\neq 0$
is \su{\nf-2} with $\nf-1$ flavors.  Its
associated dynamical superpotential
\eqn\dynsupdnftwo{
\mu^{-1}\h{M}\h{N} - \frac{\mu\det\h{N}}{\LbD M_1^1} -
{\h{B}\h{N}\h{\ot B}\over \mu M_1^1}
}
leads to \SBconstr\ through $\del{W}/\del{N_2^2}$ and
to \SBvev\ when $\h{m}\h{M}$ is added to the superpotential.

Now we wish to rotate $\ot{Q}_1\leftrightarrow Q^2$ so that
$B^{12}(Q)$ has a vacuum expectation value.  The flavor-rotated
version of \SBconstr\ is
\eqn\SBconstrB{
M_u^1(Q) B^{2r}(Q) + M_u^2(Q) B^{r1}(Q) + M_u^r(Q) B^{12}(Q) = 0\ .
}
In the magnetic \su{\nf-2} theory we have
\eqn\flrotd{
\vev{B^{12}(q)} \propto \vev{q_3 q_4 \cdots q_{\nf-2}} \neq 0\ 
}
which completely breaks the \su{\nf-2} gauge group.
If we now let $M_3^s\neq 0$ for $s=1,2,3$ we find the F-flatness
conditions
\eqn\fflatmthrr{
\vev{\aqn{3}}=0 \ ;\ \ \ \sum_{s=1}^3 \vev{M^s_3 q_s} = 0
}
Multiplying the latter equation by $\vev{q_4 \cdots q_{\nc}}$
we find \SBconstrB\ for $r=u=3$, {\it etc}.

If we add to the electric superpotential
$\ot h\ot Q_1\ot Q_2 + \bar{m}_{\bar r}^{\bar u}\fQn{\bar r}\aQn{\bar u}$,
${\bar r},{\bar u}=3,4,\dots,\nf$, we get a rotated version of
\SBvev.
\eqn\rotvers{
W_L = -\Lb \frac{\ot h\det \bar m}{\fQn{1}\fQn{2}}
}
In the magnetic theory the superpotential is
\eqn\rotversd{
\mu^{-1}M^{\bar r}_{\bar u}N^{\bar u}_{\bar r}(q)+\ot h\ot{B}_{12}(\ot q)
+ \bar{m}_{\bar r}^{\bar u}M^{\bar r}_{\bar u} \ .  } 
The F-term equations give $\fqn{\bar r}\aqn{\bar u}=-\mu\bar{m}_{\bar
r}^{\bar u}$.  Meanwhile, the D-term equations imply that $\det_{\bar
r,\bar u}\fqn{\bar r}\aqn{\bar u}=-\Lam^{-b}(-\mu)^{\nf-2}
B^{12}(q)\ot B_{12}(\ot q)$.  Substituting these relations into the
superpotential
\rotversd, one finds the first and third terms cancel while the second
term becomes
\eqn\rotversL{
\ot W_L = \ot h\ot{B}_{12}(\ot q) = -\Lb\frac{\ot h \det \bar m}{B^{12}}
}
which is the same as \rotvers.  Note that the argument is now
perturbative in the magnetic theory, in contrast to the case of meson
expectation values.

As in section 3.3,
one can again continuously connect these theories; we will not present
the details.  The key result is that the IR dynamics of the
magnetic \su{\nf-2} gauge theory with $\nf-1$ flavors is exactly
equivalent to that of a completely broken \su{\nf-2} gauge theory with
a certain superpotential, and that an infinite
class of theories interpolates between them.

\subsec{\su{2}\ with $\nf=4$ and its dual}

This theory is special, as it is dual to a theory with the same gauge
group and number of doublet representations; however the magnetic
theory contains extra singlets and a superpotential \wdual\ which breaks
its flavor symmetry from $SU(8)$ to $SU(4)\times SU(4)\times U(1)$.
If we give mass to any pair of doublets, so that six remain, we
immediately generate a dynamical superpotential \Seibone.  We will see
interesting physics from this below.

We begin by reviewing the theory with an ordinary mass term
$m\fQn{1}\aQn{1}$.  At scales below this mass, we integrate out the
massive doublets and obtain a theory which confines and generates a
dynamical superpotential:
\eqn\twofourmorig{
W_L={1\over m\Lam^2}\left(\h B\cdot\h M\cdot 
\h{\tilde B}-\det \h M\right).
}
In the magnetic theory, $m$ leads to symmetry breaking;
in this case, the superpotential is generated \NAD\ partly at
tree level, and partly by instantons in the broken group. The tree
level part is obtained by expanding around the vacuum expectation
values:
\eqn\vacexpexp{ 
\tilde W=\mu^{-1}M^r_u\fqn{r}\aqn{u}+m M^1_1\;\;\rightarrow \;\;
\tilde W_{L,tree}=\mu^{-1}M^{\h r}_{\h u}(\fqn{2})_{\h r}(\aqn{2})^{\h u}}
where $(\fqn{2})_{\h r}$ is the second color component of $\fqn{\h r}$,
and we have chosen the vevs along the first color direction so that
the first components of $\fqn{r},\aqn{u}$ are massive.
By identifying
\eqn\quarksB{ 
(\fqn{2})_{\h r}={1\over v}\sqrt{{-\mu^2\over\Lam^2}}
B^{\h s\h t}{\epsilon_{1\h r\h s\h t}\over 2!}\;\; ; \;\;\;\;\; 
(\aqn{2})^{\h u}={1\over v}\sqrt{{-\mu^2\over\Lam^2}}
\tilde B_{\h v\h w}{\epsilon^{1\h u\h v\h w}\over 2!}
} 
where $v^2=-m\mu$, we may rewrite 
this superpotential as
\eqn\twofourmtree{
\tilde W_{L,tree}={1\over m\Lam^{2}} \
{1\over 2!}\epsilon_{\h r\h s\h t}B^{\h s\h t}\
M^{\h r}_{\h u} \
{1\over 2!}\epsilon^{\h u\h v\h w}\tilde B_{\h v\h w}.
}
The remaining term comes from instantons in the unbroken group; we can
uncover this term by turning on vevs for $\h M$.\NAD\ This will give mass
to three of the four flavors; the low-energy theory in this case
contains an instanton superpotential: \ADS
\eqn\twofourmdyn{
\tilde W_{L,dyn}={\ot\Lambda_{L'}^5\over\fqn{1}\aqn{1}} =
-\ot\Lambda^2{\det \h M\over m\mu^4} =-{\det\h M\over \Lam^2 m} \ .} 
Thus the terms
\twofourmtree\ and \twofourmdyn\ combine to reproduce
Eq.~\twofourmorig\ for the low-energy magnetic theory.

Consider now the theory with
$\wtree=\lam\fQn{1}\fQn{2}$. The F-term equations tell us that
$M^1_u=M^2_u=B^{1r}=B^{2r}=0$. Integrating out the massive quarks
gives us a superpotential
\eqn\twofourhorig{
W_L={1\over\lam\Lam^2}\left( \Mitim M^s_v-{1\over 4}
B^{rs}\tilde {B}_{uv}\right) \tilde{B}_{wx}{\eps^{uvwx}\over 2!}
{\eps_{rs}\over2!}
}
which is just an \su{8}\ rotation of Eq.~\twofourmorig.

In the magnetic theory, we have a superpotential
\eqn\wdualtwofour{
\ot W_{\rm tree}=\sqrt{-\Lam^2\over\mu^2}h\fqn{3}\fqn{4}
+\mu^{-1}\sum_{a=1,2}\fqn{a}M^a_u\aqn{u}
+\mu^{-1}\sum_{i=3,4}\fqn{i}M^i_u\aqn{u}.
}
We see that $\lam\neq0$, in contrast to the case for $m\neq 0$,
does not lead to symmetry breaking;
rather, for large $\lam$, $\fqn{3},\fqn{4}$ are
massive. Integrating them out, we find
a superpotential
\eqn\twofourhtree{
\ot W_{L,{\rm 
tree}}=-\sqrt{-\mu^2\over\Lam^2}
{1\over\lam\mu^2}\aqn{u}M^3_uM^4_v\aqn{v}+\mu^{-1}\fqn{a}M^a_u\aqn{u}
}
where $a=1,2$ and $u,v=1,\ldots,4$.
Now because this theory has only three flavors, it confines,
leading to a superpotential:
\eqn\thrflavsup{\eqalign{
\ot W_L=+&{1\over\lam\Lam^2}{\eps_{rs}\over2!}{\eps^{uvwx}\over 2!}
\Mitim M^s_v\tilde{B}_{wx}+\mu^{-1}N_a^uM^a_u\cr
&-{\mu^2\over\lam\Lam^2 \ot\Lam^2}
\left(
\ot{B}_{wx}{N}^w_a {N}^x_b+
{\mu^2\over4\Lam^2}B_{ab}\ot{B}_{uv}\ot{B}_{wx}{\eps^{uvwx}\over 2!}
\right){\eps^{ab}\over 2!}
\cr}}
The first two terms are those of Eq.~\twofourhtree\ and
the last is generated dynamically; using $\Lam^2\LamD^2=\mu^4$ one 
sees that it is again merely an \su{8}-flavor
rotation of the analogous term \twofourmorig,
similar to that employed in writing Eq.~\twofourhorig.
{}From the F-term equations, we see that we must have
$\vev{M^a_u}=\vev{B^{ar}}=0$, in
agreement with the electric theory; also we have $\vev{N}=0$,
so that \twofourhorig\ and \thrflavsup\ agree.

\subsec{\su{2}\ with $\nf=5$ and its dual}

As discussed in \NAD, the superpotential $mQ^1\aQn{1}$ causes this theory
to flow to \suii\ with four flavors,  which flows to
an interacting fixed point in the IR.
The flavor symmetries ensure that a term $hQ^1\fQn{2}$ will have the
same effect.  In the magnetic $SU(3)$ gauge theory this means that the
superpotential $W\propto q_3q_4q_5$, which makes the theory chiral,
still leads to a theory whose dynamics we know.

Let us examine the magnetic theory in some detail. It has gauge group
$SU(3)$ and a superpotential
$$W=h\sqrt{\Lam\over\mu^3} q_3q_4q_5 + \mu^{-1}M^r_u \fqn{r} \aqn{u}$$ 
which leaves a global symmetry $SU(2)\times 
SU(3)\times SU(5)\times U(1)_R$.
In the IR, we have operators $B^{\h{r}\h{s}}$,
$B^{\bar{r}\bar{s}}$, $B^{\h{r}\bar{s}}$, $\tilde B_{uv}$, $M^{\bar
r}_u$, $M^{\h{r}}_u$, $N^u_{\h{r}}$ and $N^u_{\bar{r}}$, where
$\h{r}=1,2$, $\bar{r}=3,4,5$, and $u=1,\ldots,5$.  From the F-term
equations, we find
$N^u_{\h{r}}=N^u_{\bar{r}}=B^{\h{r}\h{s}}=B^{\h{r}\bar{s}}=0$ and thus
these operators are removed in the IR.  In addition, the flat directions
associated with $M^{\h{r}}_u$  are
dynamically blocked.  If $\vev{M^{\h{r}}_u}$ has rank 1, then one flavor is 
given a mass; this leaves $SU(3)$ with 4 flavors, which confines and, as in 
Eqs.~\Wdynbloc-\cantbe,
generates a superpotential with no supersymmetric ground
state. The remaining  28 fields
$B^{\bar{r}\bar{s}}$, $M^{\bar r}_u$ and $\tilde B_{uv}$ have R-charge 1
and satisfy the F-term equations
\eqn\eomtwofive{
M^{\bar r}_u B^{\bar{s}\bar{t}}\epsilon_{\bar{r}\bar{s}\bar{t}}= 0\ ; \ \ 
M^{\bar r}_u \tilde B_{vw}\epsilon^{uvwxy} = 0 .} 
This is in agreement with the low-energy electric theory, which is
$SU(2)$ with 4 flavors; its invariants consist of 28 fields $V_{ij}$
with R-charge 1 in the antisymmetric representation of the $SU(8)$
flavor symmetry.  In the electric theory the equations \eomtwofive\
arise as classical constraints, which are unmodified in the quantum
theory.

We emphasize that the $SU(3)$ magnetic gauge theory is chiral,
but its infrared physics is fully under control.  Indeed, it 
is perhaps the simplest (somewhat trivial) example 
of a chiral theory which flows to a fixed point with
a dual description in terms of a non-chiral theory.

\newsec{Various Remarks}

In this section we note a number of related implications of our
work. 

\subsec{Other Examples of Accidental Symmetries}

Many other examples of non-perturbative accidental symmetries are
known, and are too numerous to list in their entirety.  For example,
the theories of $SU(N_c)$ which appear in
Refs.~\refs{\kutsch,\ilslist} all have these properties for $\nc=2$.
Other examples include $Spin$ group dualities of \spinduals.  If one
breaks $Spin(7)$ with $\nf+2$ spinors to $SU(3)$ with $\nf$ flavors,
the low-energy theory has an $SU(\nf)\times SU(\nf)\times U(1)$ global
symmetry.  However, the dual $SU(\nf-2)$ gauge theory, with a
symmetric tensor and $\nf+2$ fields in the antifundamental
representation, has only $SU(\nf)\times U(1)$ symmetry in the UV.

Another interesting example has recently emerged in the context
of three-dimensional N=4 gauge theories \threed.  Mirror
symmetry relates two theories, with Higgs and Coulomb branches,
hypermultiplets and vector multiplets, masses and Fayet-Iliopoulos
couplings exchanged.  In general, the mass terms appear in a larger flavor
symmetry than the Fayet-Iliopoulos couplings, and the
latter transform properly under the larger symmetry only in the IR.
The non-perturbative dynamics involved in this result should be
quite interesting.  

\subsec{Ultraviolet Irrelevant and Infrared Relevant Operators}

In the remainder of this section we focus our attention on the
special behavior of baryon operators in the vacua of $SU(N_c)$ theories at
the origin of moduli space, where classically the full gauge
group is unbroken, and where quantum mechanically one often expects
conformal field theories in the IR.

One of the common features of N=1 duality as is the presence of
operators in the IR whose dimensions are far different from their
canonical values.  The baryons in the above theories are of this type.
In general, the baryon operator which we add to the magnetic
superpotential \nftwosupd\ is of dimension $\nf-2$ in the UV, which is
perturbatively marginal for $\nf=5$ and is irrelevant (thereby making
the theory perturbatively non-renormalizable) for $\nf>5$.\foot{Recall
that the superpotential in four dimensions has dimension 3, so baryons
of dimension greater than 3 have couplings with inverse mass
dimensions and are irrelevant at weak coupling.}  However, in the IR
this baryon operator is a relevant operator.  For $\nf=3,4,5$ the
theory flows to an interacting fixed point; the superconformal algebra
tells us the baryon has dimension $1,\frac{3}{2},\frac{9}{5}$ \NAD.
For $\nf>5$ the theory flows to a free $SU(2)$ gauge theory, whose
baryons have dimension 2.

  To say this another way, one might naively have the impression from
perturbation theory that adding a high-dimension baryon to the
superpotential would not affect the IR behavior of the theory.  As we
have seen, however, such operators can indeed play a crucial role,
even changing the phase of the theory.  For example, consider the
magnetic $SU(5)$ theory with 7 flavors and mesons that flows in the IR
to a weakly-coupled electric $SU(2)$ gauge theory with $7$ flavors.
Adding to this theory's superpotential a dimension-five baryon
operator with coefficient $h^{r_1r_2}$ of rank $1$ removes one flavor
in the low-energy effective $SU(2)$ theory without changing the phase.
However, if $h$ has rank 2 then the $SU(5)$ theory hits a fixed point
at finite coupling (which is physically equivalent to the theory of
$SU(2)$ with $\nf=5$), while if it has rank 4 the theory flows to
strong coupling, confines, and has massless mesons and baryons.  Other
dramatic effects of similar operators (known as ``dangerously
irrelevant operators''; see the appendix of \kutsch\ for a brief
discussion) have been observed elsewhere
\refs{\kinsso,\kutsch,\ilslist,\emop,\DSBaryons}.  

While we so far have only explored the physics of $SU(2)$ theories
with $2\nf$ doublets and their magnetic duals, there are other
interesting \sunc\ theories with IR relevant baryons, which we will
explore further in the following sections.  Before beginning, we list
the various possibilities.

The dimensions of baryons are the same for theories with \sunc\ and
$SU(\nf-\nc)$ gauge groups.  These dimensions are determined by the R
charges of their constituents.  Since these charges depend in these
theories only on the $U(1)_R$-gauge anomalies and not on the presence
or absence of gauge singlet mesons, a theory with gauge group
$SU(\nc)$ (with a dual representation as $SU(\nf-\nc)$ with singlets)
has a baryon of the same IR dimensions as an $SU(\nf-\nc)$ theory
(with a dual representation as \sunc\ with singlets.)

In particular, theories with $SU(\nf-2)$ gauge group and {\it no}
mesons \Miti\ therefore have baryons at low-energy with the same
dimension as those with $SU(2)$ gauge group, the same number of
flavors, and {\it no} mesons \Miti.  Another easy case is $SU(\nf-1)$, for
which the dual has no gauge group; the low-energy theory always has the
baryon as a fundamental field of dimension 1 \Seibone.  In all of
these theories, addition of baryon operators to the superpotential may be
easily analyzed, since these operators are either linear or bilinear
in the fields in at least one description of the theory.

However, there are other examples where the baryons are neither
linear nor bilinear in any description.
These include $SU(3)$ for $3<\nf$, and by association $SU(\nf-3)$
theories; for $\nf>8$ the baryons have their canonical dimension (namely 3)
at the free $SU(3)$ IR gauge theory, while for $\nf=4,5,6,7,8$ the
baryons have IR dimension
$1,2,\frac{9}{4},\frac{18}{7},\frac{45}{16}$.  $SU(4)$ with $\nf=8$
has a strongly coupled fixed point with a baryon which is
IR-marginal (though not, in general, exactly marginal.)

In all other $SU(\nc)$ gauge theories with $\nf$ flavors, the
baryons are irrelevant in the IR, so adding them to the
superpotential has no effect on the conformal theory at the
origin of moduli space.

\subsec{Baryons in $SU(\nf-2)$ Gauge Theories and New Fixed Points}

Previously our focus was on non-perturbative accidental
symmetries, so we have considered as the electric theory only the
theory of $SU(2)$ with $2\nf$ doublets.  But we may also consider the
physics of $SU(\nf-2)$ with $\nf$ flavors and no singlets as the
electric theory.  In this case, both this theory and its dual have the
usual $SU(\nf)_L\times SU(\nf)_R\times U(1)_B$ flavor symmetry and
there are no accidental IR symmetries.  Nevertheless, the baryon
operators are dual to mass terms in the $SU(2)$ magnetic theory and
one can analyze these theories in some detail.

In particular, one is immediately led to a number of new fixed points,
some of which are certainly present, others of which are likely to be
present.  We noted in a previous section that $SU(2)$ with 10
doublets, two of which are massive, is dual to an $SU(3)$ magnetic
theory with $\nf=5$ and a single baryon operator in its
superpotential; both theories flow in the IR to a fixed point
characteristic of $SU(2)$ with 8 doublets.  We may now consider the
behavior of $SU(3)$ with $\nf=5$ as the electric theory.  The effect
of a superpotential $W=hQ^1Q^2Q^3$ is to change the dual
superpotential to $\tilde
W=\mu^{-1}(\fqn{r}\Mitim\aqn{u}+h\Lam^2q_4q_5)$.  Integrating out the
massive fields $q_4, q_5$ leaves
\eqn\suthreed{
\tilde W = \sum_{r=1}^3\sum_{u=1}^5 \mu^{-1}\fqn{r}\Mitim\aqn{u}
} 
plus terms proportional to $1/h$ that are irrelevant in the IR.  The
IR behavior of this chiral $SU(2)$ gauge theory is not known but it is
likely that it flows to a fixed point.  The properties of this fixed
point are not exactly determined; the symmetries are broken enough
that the R charges of the $q$ and $\tilde q$ fields, and therefore
their dimensions, cannot be determined independently.

More generally, if we consider \su{\nf-2} with $\nf$ flavors
as the electric theory with $W=hB+\tilde h\tilde B$, with
$h$ of rank $p$ and $\tilde h$ of rank $\tilde p$, the
dual theory is $SU(2)$ with $2(\nf-p-\tilde p)$ doublets
and $(\nf-2p)(\nf-2\tilde p)$ mesons \Miti\ coupled as
\eqn\sunfmintwod{
\tilde W = \sum_{r=1}^{\nf-2p}\sum_{u=1}^{\nf-2\tilde p} 
\mu^{-1}\fqn{r}\Mitim\aqn{u}
} 
For $2(\nf-p-\tilde p)\geq 12$, these theories are free in the IR and
their behavior is therefore straightforward.  For $2(\nf-p-\tilde p)=
8$ or $10$ the situation is more interesting.  If $p=\tilde p$,
$\nf-2p=0$, or $\nf-2\tilde p=0$, these theories are already known
\NAD\ to reach interacting fixed points, at which the dimensions of
all chiral operators can be determined.  For other cases the theories
have not been considered.  However it seems plausible that all of them
reach new, previously unidentified interacting fixed points.  Note
that these theories are chiral and the global symmetries are
insufficient to determine the dimensions of most chiral operators.
For $2(\nf-p-\tilde p)<8$ these theories probably confine and 
generate dynamical superpotentials, which we have not analyzed.

\subsec{Other New Fixed Points Associated with Baryon Perturbations}

For those theories with IR-relevant baryon operators which are not
dual to mass terms, it is impossible to analyze their non-perturbative
effects in detail.  Nonetheless, as they certainly cause the theory to
flow away from its original fixed point, it is worth speculating about
the physics of this process.  For theories with IR-marginal baryon
operators, the question is whether the operators are exactly marginal,
marginally relevant or marginally irrelevant.  We will argue that in
most cases they are expected to be marginally irrelevant, but, as
shown in \emop, in two independent cases they are exactly marginal.

Let us first consider theories with $SU(3)$ gauge group.  For
$\nf=4,5$ the discussion has already been covered.  For $\nf=6,7,8$ we
may consider the effects of adding to the superpotential $hQ^1Q^2Q^3$;
the resulting chiral theory probably flows to a previously
unidentified fixed point, in analogy to the $\nf=5$ case.  We may also
consider baryon terms of higher rank.  While there is quite a bit of
interesting physics in such an analysis, the most interesting case is
that of $\nf=6$ with $W=hQ^1Q^2Q^3 + h'Q^4Q^5Q^6$.  We may expect a
basin of attraction in which $h=h'$ in the infrared; if this theory
reaches a fixed point then it has enough symmetry that despite its
chiral nature all the R charges may be determined.  In particular, the
R charges of the six $Q^r$ are all $\frac{2}{3}$ (thus they have
dimension 1) while those of the $\tilde Q_u$ are $\frac{1}{3}$
(dimension $\frac{1}{2}$.)

For $SU(3)$ with $\nf\geq9$, the theory is not asymptotically
free and one may easily check that the baryons are perturbatively
irrelevant at weak coupling. However \refs{\suthreefinite,\emop},
there is a well-known exception for $\nf=9$.  The superpotential
\eqn\suthreeemop{
W=h(Q^1Q^2Q^3 +Q^4Q^5Q^6+Q^7Q^8Q^9+
\tilde Q_1\tilde Q_2\tilde Q_3 +\tilde Q_4\tilde Q_5\tilde Q_6
+\tilde Q_7\tilde Q_8\tilde Q_9)
}
combined with a non-zero gauge coupling $g$
actually represents an exactly marginal perturbation of the
free fixed point; that is, there is a complex curve $F(g,h)=0$ passing
through $g=h=0$ where this theory is conformal.  One may
see hints of this by looking in first-order perturbation theory; however
this result holds to all orders \allorders\
and in fact holds non-perturbatively \emop.

We omit the discussion of $SU(\nf-3)$, since the analysis
is almost identical to that of  $SU(3)$ theories with $\nf$ flavors.
We merely note that $SU(6)$ with $\nf=9$ has an exactly marginal operator
which is dual to that of Eq.~\suthreeemop\ \emop.

Finally we consider $SU(4)$ with $\nf=8$.  This theory has a
non-trivial IR fixed point where its baryon becomes marginal.  However
it is certainly not exactly marginal (and is probably marginally
irrelevant, in analogy to the case of $SU(3)$ with $\nf=9$ considered
above) unless one considers a superpotential of high symmetry, in
analogy to \suthreeemop\ \emop
\eqn\sufoureemop{
W=h(Q^1Q^2Q^3Q^4+Q^5Q^6Q^7Q^8+
\tilde Q_1\tilde Q_2\tilde Q_3\tilde Q_4+\tilde Q_5\tilde Q_6
\tilde Q_7\tilde Q_8) \ .
}
This operator (combined with a variation in the gauge coupling)
is exactly marginal, even though it is irrelevant in the UV
and makes the theory naively non-renormalizable.
 
\newsec{Conclusion}

We may summarize our main result formally as follows.  It may happen
that while a theory $T_E$ has a global symmetry $G_E$ in the UV which
is preserved in the IR, its magnetic dual $T_M$ has a smaller global
symmetry $G_M$ in the UV, and this global symmetry is enhanced to
$G_E$ only in the IR.  In this case the theory $T_M$ is actually only
one of a space of theories which share the same infrared behavior with
$T_E$; this space has the form $G_E/G_M$.  Trivial global rotations by
elements of $G_E$ in the electric theory are manifested as non-trivial
translations inside the space of magnetic theories $G_E/G_M$.

In the specific example studied here, the effect of the \su{2\nf}\
global symmetry of the electric \su{2}\ theory can be easily observed
by perturbing the theory.  Each perturbation generates a continuously
infinite class of magnetic duals, parametrized  by a subspace of
$SU(2\nf)/(SU(\nf)\times SU(\nf)\times U(1))$ depending on the 
perturbation.  These classes contain magnetic theories with
different gauge and flavor groups at the UV cutoff, all of which flow
to the same IR dynamics.  Unlike previously known examples of
accidental symmetries, the physics which makes this possible involves
non-perturbative dynamics.  This is a general property of duality in
the presence of accidental symmetries.

 One noteworthy aspect of this physics is that it depends on the fact
that certain baryons which are of high dimension in the UV become
relevant in the IR through the effects of strong coupling.  We have
briefly discussed additional implications of this phenomenon, including
new examples of phase transitions due to naively irrelevant operators,
new chiral fixed points, and exactly marginal operators of high canonical
dimension. 

\bigskip

\noindent {\bf Acknowledgements:} M.J.S. was supported by 
National Science Foundation grant NSF PHY-9513835 and
by the WM Keck Foundation.  
These results were obtained at Rutgers University under the support of
DOE grant \#DE-FG05-90ER40559.
\

\listrefs
\end